\let\accentvec\vec 
\documentclass{llncs}
\let\vec\accentvec %

\usepackage[utf8]{inputenc}
\usepackage[T1]{fontenc} 

\usepackage{amsmath}
\usepackage{amsfonts}
\usepackage{amssymb}

\usepackage[ruled,vlined,linesnumberedhidden]{algorithm2e}
\DontPrintSemicolon

\usepackage{graphicx}
\usepackage{subfigure}
\usepackage{xcolor}
\usepackage{tikz}

\definecolor{darkblue}{rgb}{0,0,.5}
\definecolor{darkred}{rgb}{0.5,0,0}
\definecolor{lightgray}{rgb}{0.8,0.8,0.8}
\definecolor{lightlightgray}{rgb}{0.9,0.9,0.9}

\usepackage{paralist}
\usepackage{colortbl}
\usepackage{placeins} 
\usepackage{framed}
\usepackage{wrapfig}

\usepackage{url}
\usepackage{hyperref}
\hypersetup{colorlinks=true, breaklinks=true, linkcolor=darkred, menucolor=darkred, urlcolor=darkred, citecolor=darkred}

\newcommand{\cost}{\gamma}
\newcommand{\mcost}{\hat{\gamma}}
\newcommand{\nprim}{{n_{\mathrm{p}}}}
\newcommand{\nback}{{n_{\mathrm{b}}}}
\newcommand{\fr}{\lambda}
\newcommand{\npn}{r_p}
\newcommand{\Enpn}{\overline{r_p}}
\newcommand{\npm}{\alpha_p}
\newcommand{\Enpm}{\overline{\alpha_p}}
\newcommand{\pr}{{\scriptstyle \#}\mathrm{pr}}
\newcommand{\Epr}{\overline{{\scriptstyle \#}\mathrm{pr}}}
\newcommand{\Vpr}{S^2[{\scriptstyle \#}\mathrm{pr}]}
\newcommand{\st}{{\scriptstyle \#}\mathrm{st}}
\newcommand{\Est}{\overline{{\scriptstyle \#}\mathrm{st}}}
\renewcommand{\Vst}{S^2[{\scriptstyle \#}\mathrm{st}]}
\newcommand{\res}{\sum_{\mathrm{res}}}
\newcommand{\kprim}{{k_{\mathrm{p}}}}
\newcommand{\kback}{{k_{\mathrm{b}}}}
\newcommand{\gc}{\mathrm{globalCounter}}
\newcommand{\palpha}{\mathfrak{a}}
\newcommand{\pbeta}{\mathfrak{b}}
\newcommand{\wb}{w}
\newcommand{\Ewb}{\overline{w}}
\newcommand{\Vwb}{S^2[w]}
\newcommand{\tnb}{{c^{\scriptscriptstyle-}_{s,m}}}
\newcommand{\tb}{{c^{\scriptscriptstyle+}_{s,m}}}

\newcommand{\cs}{c^{\scriptstyle\mathrm{start}}}
\newcommand{\ce}{c^{\scriptstyle\mathrm{end}}}

\newcommand{\ie}{i.e.}
\newcommand{\eg}{e.g.}

\newcommand{\imgScale}{0.58}
\newcommand{\tabScale}{0.75}
\newcommand{\figPath}{.}

\spnewtheorem*{prob}{Problem}{\itshape}{\upshape}

\author{%
Martin Dietzfelbinger\inst{1}\thanks{Research supported by DFG grant DI 412/10-1.}
\and %
Michael Mitzenmacher \inst{2}\thanks{Research supported by NSF grants IIS-0964473 and CCF-0915922.}
\and %
Michael Rink\inst{1}${}^\star$
}

\institute{%
Fakult\"at f\"ur Informatik und Automatisierung, Technische Universit\"at Ilmenau
\email{\{martin.dietzfelbinger,michael.rink\}@tu-ilmenau.de}
\and %
School of Engineering and Applied Sciences, Harvard University
\email{michaelm@eecs.harvard.edu}
}

\title{Cuckoo Hashing with Pages}

\begin{document}
\maketitle
\begin{abstract}  
Although cuckoo hashing has significant applications in both
theoretical and practical settings, a relevant downside is that it
requires lookups to multiple locations.  In many settings, where
lookups are expensive, cuckoo hashing becomes a less compelling
alternative.  One such standard setting is when memory is arranged in
large pages, and a major cost is the number of page accesses.  We
propose the study of cuckoo hashing with pages, advocating approaches
where each key has several possible locations, or cells, on a single
page, and additional choices on a second backup page.  We show
experimentally that with $k$ cell choices on one page and a single
backup cell choice, one can achieve nearly the same loads as when each
key has $k+1$ random cells to choose from, with most lookups requiring
just one page access, even when keys are placed online using 
a simple algorithm.  While our results are currently
experimental, they suggest several interesting new open theoretical
questions for cuckoo hashing with pages.
\end{abstract}  
\section{Introduction}
Standard cuckoo hashing places keys into a hash table by providing
each key with $k$ cells determined by hash functions.  Each cell can
hold one key, and each key must be located in one of its cells.  As
new keys are inserted, keys may have to move from one alternative to
another to make room for the new key. Cuckoo hashing provides high
space utilization and worst-case constant-time lookups, making it an
attractive hashing variant, with useful applications in both
theoretical and practical settings, \eg, \cite{alcantara2009real,erlingsson2006cool,FotakisPSS2005,PaghR2004,ross2007efficient}.  

Perhaps the most significant downside of cuckoo hashing, however, is
that it potentially requires checking multiple cells randomly
distributed throughout the table.  In many settings, such random
access lookups are expensive, making cuckoo hashing a less compelling
alternative.  As a comparison, standard linear probing works
well for many settings where memory is split into (not too small)
chunks, such as cache lines; in such settings, with suitably small
loads, the average number of memory accesses is usually very close to
1.

In this paper, we consider cuckoo hashing under a setting where memory
is arranged into pages, and the primary cost is the number of
page accesses.  In such a setting, a natural scheme to minimize this
number might be to first hash each key to a page, and
then keep a separate cuckoo hash table in each page.  This limits the
number of pages examined to one, and maintains the constant lookup
time once the page is loaded.  Such a scheme has been utilized in
previous work (\eg, \cite{alcantara2009real}).  However, a problem
with such a scheme is that the most overloaded page limits the load
utilization of the entire table.  As we show later, the random
fluctuations in the distribution of keys per page can significantly
affect the maximum achievable load.

We generalize the above approach by placing most of the cell choices
associated with a key on the same primary page.  We then allow a
backup page to contain secondary choices of possible locations for a
key (usually just one).  In the worst case we now must access two
pages, but we demonstrate experimentally that we can arrange so that
for most keys we only access the primary page, leading to close to one page
accesses on average.  Intuitively, the secondary page for each key
allows overloaded pages to slough off load constructively to
underloaded pages, this distributing the load.  We show that we
can do this effectively offline as well as online by evaluating an
algorithm that we find performs well even when keys are deleted as
well as inserted into the table.

We note that it is simple to show that using a pure splitting scheme,
with no backup page, and page sizes $s=m^\delta$, $0<\delta<1$, where
$m$ is the number of memory cells, the load thresholds obtained are
asymptotically provably the same as for cuckoo hashing without pages.
Analysis using such a parametrization does not seem suitable to
describe real-world page and memory sizes.  While we \emph{conjecture}
that the load thresholds obtained using the backup approach, for
reasonable parameters for memory and page sizes, match this bound, at
this point our \emph{work is entirely experimental}.  We believe this
work introduces interesting new theoretical problems for cuckoo
hashing that merit further study.
\subsection{Related Work}
The issue of coping with pages for hash-based data structures is not
new.  An early reference is the work of Manber and Wu, who consider
the effects of pages for Bloom filters
\cite{manber1994algorithm}. Their approach is the simple splitting approach we
described above; they first hash a key to a page, and each page then
corresponds to a separate Bloom filter.  The deviations in the number
of keys hashed to a page yield only small increases in the overall
probability of a false positive for the Bloom filter, making this
approach effective.  As we show below, such deviations
have more significant effects on the acceptable load for cuckoo
hash tables, leading to our suggested use of backup pages.  More recent
work includes that of Woelfel \cite{woelfel2006maintaining}, who
focuses on perfect external memory dictionaries that require
additional space for the hash function and for handling insertions and
deletions.

Our work is perhaps superficially related to the body of literature on
cuckoo hashing where cells (or buckets) can hold multiple keys. 
Here for searching the whole page or bucket has to be scanned.
This topic was first examined by Dietzfelbinger and Weidling
\cite{dietzfelbinger2007balanced}; other notable work in this area
includes that of Lehman and Panigrahy, who prove that ``overlapping''
buckets can yield improved space utilization \cite{LehmanP2009}.  Here
we don't think of the entire page as a bucket, as we are thinking of
pages as being sufficiently large that we may want to avoid searching
through a page for a key.  Our work can also be considered as related
to work on using stashes, or extra locations when keys cannot be
placed normally, with cuckoo hashing \cite{kirsch2008more}.  Here,
each key can be thought of as having an individualized stash corresponding
to one or more cells on a separate page.

A number of papers have recently resolved the longstanding issue regarding the load
threshold for standard cuckoo hash tables where each key obtains $k$
choices
\cite{dietzfelbinger2010tight,fountoulakis2011multiple,fountoulakis2010orientability,frieze2009maximum,GaoWormald2010}.
Our work re-opens the issue, as we consider the question of the effect
of pages on these thresholds, if the pages are smaller then
$m^\delta$, such as for example $\mathrm{polylog}(m)$.

Practical motivation for this approach includes recent work on
real-world implementations of cuckoo hashing \cite{alcantara2009real,ross2007efficient}.
In \cite{alcantara2009real}, where cuckoo
hashing algorithms are implemented on graphical processing units, the
question of how to maintain page-level locality for cuckoo hash tables
arises.  Even though work for lookups can be done in parallel, the
overall communication bandwidth can be a limitation in this setting.
Ross examines cuckoo hashing on modern processors, showing they
can be quite effective by taking advantage of available parallelism
for accessing cache lines \cite{ross2007efficient}.  Our approach 
can be seen as attempting to extend this performance, from cache lines
to pages, by minimizing the amount of additional parallelism required.
\subsection{Our Results}
We give a short summary of the results in the paper. (All results are
experimental.)  Our presented results focus on the setting of four
location choices per key.  The maximum load factor $c^*_4$ of keys
with four hash functions and no paging is known.  With small pages and
each key confined to one page, we find using an optimal offline
algorithm that the maximum achievable load factor is quite low, well
below $c^*_4$.  However, if each key is given three choices on a
primary page and a fourth on a backup page, the load factor is quite
close to $c^*_4$, even while
placing most keys in their primary page, so that most keys can be
accessed with a single page access.  With three primary choices, a
single backup choice and filling up the table to 95 percent,
we find that only about 3 percent of keys need
to be placed on a backup page (with suitable page sizes).  We show
that a simple variation of the well-known random walk insertion
procedure allows nearly the same level of performance with online,
dynamic placement of keys (including scenarios with alternating insertion and deletions).
Our experiments consistently yield that at
most 5 percent of keys needs to be placed on a backup page with these
parameters. This provides a tremendous reduction of the number of
page accesses required for successful searches. 
For unsuccessful searches, spending a little more space for
Bloom filters on each page leads to an even smaller number of accesses to backup pages.

\section{Problem Description}

We want to place $n$ keys into $m=n/c$ memory (table) cells where each cell can hold a fixed number of $\ell\geq 1$ keys.  The value $c$ is referred to 
as the load factor.  
The memory is subdivided into $t$ pages (sub-tables) of equal size $s=m/t$.  (Throughout the paper we assume
$m$ is divisible by $t$.)  Each key is associated with a \emph{primary page} 
and a \emph{backup page} distinct from the primary page, as well as a set of $k$ distinct table cells, $\kprim$ on the primary page and $\kback=k-\kprim$ on the backup page.  The pages and keys are chosen according to hash functions on the key, and it is useful to think of them as being chosen uniformly at random in each case. 
For a given assignment let $\nprim$ be the number of keys that are placed in their primary page and let $\nback$ be the number of keys that are placed in their backup page. We can state the \emph{cuckoo paging problem as} follows.

\begin{prob}[Cuckoo Paging]
Find a placement of the $n$ keys such that the fraction ${\nprim}/{n}$ is maximized.
\end{prob}
\begin{remark}
Note that under the standard model, with no backup pages and all key locations
assumed to be chosen uniformly at random, there is threshold load factor $c^*_{k,\ell}$ such that whenever $c < c^*_{k,\ell}$ a placement exists with probability $1-o(1)$. The recent paper \cite{fountoulakis2011multiple} gives the complete picture for all reasonable values of $k$ and $\ell$.
\end{remark}

\begin{remark}
As mentioned, if $\kback=0$ and page sizes are $s=m^\delta, \delta>0$,
the asymptotic threshold load factor is the same as in the setting without pages.
This is easily proven using tight concentration bounds on the number of keys per page.
Our interest, however, is in ranges for $m$ that are realistic and not too
large page sizes $s$, so that this asymptotic behavior is not an adequate description of performance.
\end{remark}

The aim of the paper is to experimentally investigate the potential
for saving access cost by using primary and backup pages. Appropriate
algorithms are presented in the next section. For ease of description
of the algorithms we also use the following bipartite cuckoo graph model as well as the
hashing model.


\subsection{Cuckoo Graph Model}
\label{sec:graphModel}
We consider random bipartite graphs $G=(L \cup R ,E )$ with left node set $L=[n]$ and right node set $R=[m]$.
The left nodes correspond to the keys, the right nodes correspond to the memory cells of capacity $\ell$. The set $R$ is subdivided into $t$ segments $R_0,R_1,\ldots,R_{t-1}$, each of size $s=m/t$, which correspond to the separate pages.
Each left node $x$ is incident to $k=\kprim+\kback$ edges where its neighborhood $N(x)$ consists of two disjoint sets $N_p(x)$ and $N_b(x)$ determined according to the following scheme (all choices are fully random): choose $p$ from $[t]$ (the index of the primary page); then choose $\kprim$ different right nodes from $R_p$ to build the set $N_p(x)$; next choose $b$ (the index of the backup page) from $[t]-\{p\}$; and finally choose $\kback$ different right nodes from $R_b$ to build the set $N_b(x)$. Let $e=\{x,y\}$ be an edge where $x\in L$ and $y\in R$. We call $e$ a \emph{primary edge} if $y \in N_p(x)$ and call $e$ a \emph{backup edge} if $y \in N_b(x)$.

\section{Algorithms}
Using the cuckoo graph we can restate the problem of inserting the keys as finding an \emph{orientation} of the edge set of the cuckoo graph $G$ such that the indegree of all left nodes is exactly 1 and the outdegree of all right nodes is at most $\ell$.
We call such an orientation \emph{legal}. An edge $e=(y,x)$ with $x$ from $L$ and $y$ from $R$ is interpreted as ``storing key $x$ in memory cell $y$.'' If $y$ is from $N_p(x)$ we call $x$ a \emph{primary key} and otherwise we call $x$ a \emph{backup key}.
Each legal orientation which has a maximum number of primary keys is called \emph{optimal}.

\subsection{Static Case}
In the static case, \ie, if the cuckoo graph $G$ is given in advance, 
there are well-known efficient (but not linear time) algorithms to find an optimal orientation of $G$. One possibility is 
to consider a corresponding \emph{minimum cost matching problem}:
Assign costs to each edge from $G$ where primary edges get cost $0$ and backup edges get cost $1$.
Then replace each node $y$ from $R$ with $\ell$ copies and replace each edge to which $y$ is incident with $\ell$ copies as well.
Initially direct all edges from left to right. Edges from right to left are \emph{matching edges}. The minimum cost matching problem is to find a left-perfect matching (legal orientation) with minimum cost (minimum number of backup keys).
The algorithm we used to determine such a matching is a variant of the Successive Shortest Path Algorithm \cite{network_flows} but uses a modified Hopcroft-Karp Algorithm instead of Dijkstra's Algorithm for finding augmenting paths of minimal cost. 

Given a bipartite graph with $0$-$1$ edge costs the modified Hopcroft-Karp Algorithm finds a left-maximum matching of minimum cost
as follows.
Initially let $\mcost=0$. The algorithm works in rounds. In each round we try to find node disjoint augmenting paths (directed paths with free start node from $L$ and free end node from $R$) of cost exactly $\mcost$. Consider round number $i$.
For each augmenting path found in round $i$ flip the edge orientations and the edge costs along the path, and then go to round $i+1$. If in  round $i$ no such path exists but there is an augmenting path of larger costs, increment $\mcost$ by one and go to round $i+1$; otherwise stop the algorithm. Augmenting paths with fixed costs $\mcost$ are found via a combination of a modified breadth first search (BFS) and depth first search (DFS). 
The BFS starts from all left nodes with in-degree zero $L_0$ (unmatched nodes) at the beginning of a round.
The search partitions the nodes into layers. For each explored node the layer and the costs of the path to this node are stored.
A node can be explored twice if it is reached by a path of lesser cost. The BFS stops at the first level where one or more free nodes of $R$ are reached by a path of cost exactly $\mcost$. Let $R_0$ be the set of these right nodes. The algorithm tries to find node disjoint path between $R_0$ and $L_0$ via DFS, where the search can only follow edges between two successive layers.
During the recursive descent the costs of the path are accumulated. If the DFS reaches a free node and the costs are exactly $\mcost$,
the recursive ascent removes the node labels, flips the edge costs and orientations along the path.
The algorithm is optimal in the sense that it finds a left-maximum matching of minimum costs since we have only integer weights and the costs of the minimum cost augmenting paths are monotonically non-decreasing.

\begin{algorithm}[!ht] 
\small
\caption{\texttt{ModifiedHopcroftKarp(\textsf{bipartite\_graph} $G$)}\label{algo:modHK}}

\SetKw{KwFalse}{false}
\SetKw{KwNot}{not}

$\mcost \leftarrow 0$\;
\While{\upshape{an augmenting path exists}}
{
$L_0\leftarrow$ $\{x \mid x\in L \text{, $x$ unmatched}\}$\;
$R_0\leftarrow$ \texttt{BFS}($\mcost$)\;
\textrm{atLeastOnePathFound}$\leftarrow$ \KwFalse\;
    \ForEach{$y \in R_0$}
    {
        \textrm{atLeastOnePathFound}$\leftarrow $\texttt{DFS}$(y,0,\mcost)$\;
    }
    \lIf{\KwNot \upshape{\textrm{atLeastOnePathFound}}}
    {
        $\mcost\leftarrow \mcost+1$\;
    }
}
\texttt{BFS}(\textsf{max\_cost} $\mcost$):
\begin{compactitem}
 \item partitions the nodes into layers, starting from $L_0$
 \item stops at the first layer $l$ with path to a right node with cost equals $\mcost$
 \item returns the set of free nodes at layer $l$ with path cost equals $\mcost$
\end{compactitem}
 
\texttt{DFS}(\textsf{node} $y$, \textsf{current\_cost} $\cost$, \textsf{max\_cost} $\mcost$):
\begin{compactitem}
 \item recursive descent through the layers given by \texttt{BFS}
 \item the current costs of the path are accumulated in $\cost$
 \item if a free node  is reached and $\cost=\mcost$ then the recursive ascent removes the node labels, flips the edge costs and orientations along the path and returns true; otherwise returns false
\end{compactitem}

\end{algorithm}

\subsection{Dynamic Case}
\label{sec:randomWalk}
In the online scenario the cuckoo graph initially consists only of the right nodes. To begin let us consider the case of insertions only.  The keys arrive and are inserted one by one,
and with each new key the graph grows by one left node and $k$ edges. To find an appropriate orientation of the edges in each insertion step, we use a random walk algorithm, which is a modification of the common random walk for $k$-ary cuckoo hashing \cite{FotakisPSS2005} but with two additional constraints:
\begin{compactenum}
 \item avoid creating backup keys at the beginning of the insertion process, and
 \item keep the number of backup keys below a small fixed fraction.
\end{compactenum}
For the description of the algorithm we use a dual approach. The pseudocode (Algorithm \ref{algo:random_walk})
refers to the graph model and the following explanation uses the hashing model.
We refer to a key's $\kprim$ cells on its primary page as primary positions, and the
$\kback$ cells on its backup page as backup positions.
The insertion of an arbitrary key $x$ takes one or more basic steps of the random walk,
which can be separated into the following sub-steps.
\begin{algorithm}[!ht] 

\small
\caption{\texttt{RandomWalkInsert(\textsf{node} $x$)}\label{algo:random_walk}}

\SetKw{KwTrue}{true}
\SetKw{KwFalse}{false}
\SetKw{KwAnd}{and}
\SetKw{KwNot}{not}
\SetKw{KwBreak}{break}

\newcommand{\od}{\mathrm{outdeg}}
\newcommand{\suc}{\mathrm{success}}

\newcommand{\rn}{\texttt{randomNumber()}}


$\suc\leftarrow \KwFalse$\;
\While{$\gc>0 $ \KwAnd \KwNot $\suc$}
{
    \If{$\exists \ y \in N_p(x)$ with $\od(y)<\ell$}
    {
            flip edge $(x,y)$; $\suc\leftarrow \KwTrue$\;
        
    }

    \If{\KwNot $\suc$}
    {
        \eIf{ \emph{\rn}$<\palpha$}
        {
                choose random $y \in N_p(x)$;
                choose random $x'\in N(y)$\;
                flip edge $(x,y)$; flip edge $(y,x')$; $x\leftarrow x'$\;
        }
        {
            \eIf{$\exists \ y \in N_b(x)$ with $\od(y)<\ell$}
            {
                flip edge $(x,y)$; $\suc\leftarrow \KwTrue$\;
            }
            {
                choose random $y \in N_b(x)$;
                choose random $x'\in N(y)$\;
                flip edge $(x,y)$; flip edge $(y,x')$; $x\leftarrow x'$\;
            }
        }
    }
$\gc\leftarrow \gc-1$\;
}
\KwRet $\suc$\;
{\scriptsize \textrm{\texttt{(*}The modification to avoid unnecessary back steps is not shown for the sake of clarity.\texttt{*)}}}
\end{algorithm}

Let $x$ be the key that is currently ``nestless'', \ie, $x$ is not stored in the memory.
First check if one of its primary positions is free. If this is the case store $x$ in such a free cell and stop successfully. Otherwise toss a biased coin to decide whether the insertion of $x$ should be proceed on its primary page or on its backup page.
\begin{compactitem}
\item If the insertion of $x$ is restricted to the primary page, randomly choose one of its primary positions $y$.
Let $x'$ be the key which is stored in cell $y$. Store $x$ in $y$, replace $x$ with $x'$, and start the next step of the random walk.
\item  If $x$ is to be stored on its backup page, first check if 
one of the backup positions of $x$ is free. If this is the case store $x$ in such a free cell and stop successfully. Otherwise randomly choose one of the backup positions $y$ on this page and proceed as in the previous case.
\end{compactitem}

The matching procedure is slightly modified to avoid unnecessary back steps. That is, if a key $x$ displaces a key $x'$ and in the next step $x'$ displaces $x''$ then $x''=x$ is forbidden as long as $x'$ has another option on this page.

The algorithm uses two parameters. 
 \begin{compactitem}
 \item [$\palpha$] - the bias of the virtual coin. This influences the fraction of backup keys. 
 \item [$\pbeta$]  - controls the terminating condition. A global counter is initialized with value $\pbeta\cdot n$, which is the maximum number of total steps of the random walk summed over all keys. For each basic step the global counter is decremented by one. If the limit is exceeded the algorithm stops with ``failure''.
 \end{compactitem}

Deletions are carried out in a straightforward fashion. To remove a
key $x$, first the primary page is checked for $x$ in its possible
cells, and if needed the backup page can then be checked as well.  The
cell containing $x$ is marked as empty, which can be interpreted as
removing the left node $x$ and its $k$ incident edges from $G$. The
global counter is ignored in this setting ($\pbeta = \infty$).

\section{Experiments}
For each of the following experiments we consider cuckoo graphs $G$ randomly generated according to some configuration 
$\kappa=(c,m,s, \kprim, \kback)$ where $c$ is the quotient of left nodes (keys) and right nodes (table cells), $m$ is the total number of right nodes, $s$ is the page size, and $\kprim,\kback$ are the number of primary and backup edges of each left node.
In the implementation the left and right nodes were simply the number sets $[n]$ and $[m]$. All random choices were made via
the pseudo random number generator {$\mathrm{MT}19937$} ``Mersenne Twister'' of the GNU Scientific Library~\cite{GNU_Scientific}.

If not stated otherwise the total number of cells is $m=10^6$ and pages are of size $s=10^i, i\leq 6$. Our main focus is on situations where $\ell=1$, \ie, each cell can hold one key. Moreover we restrict ourselves to the cases $\kprim=3, \kback=1$ and (just for comparison) $\kprim=4$ and $\kback=0$.  While we have done experiments with other parameter values, we believe these settings portray the main points.  Also, while we have computed sample variances, in many cases they are small;  this should be assumed when they are not discussed.  

\subsection{Static Case}
Experimental results for the static case determine the limits of our approach and serve as a basis of comparison for the dynamic case.
\subsubsection{Setup and Measurements.}
First of all we want to see the limits of cuckoo hashing with pages if there are no backup options at all.
Note that for fixed page size $s$ and larger and larger table size $m$ the fraction of keys that can be placed decreases.
For $n=c\cdot m$ keys the load of each page is approximately Poisson distributed with parameter $c\cdot s$; 
asymptotically the success probability
can be estimated as
\begin{equation}
O\left(\Big(\Pr\big(\mathrm{Po}(c\cdot s) \leq s \big)\Big)^t \right)
= O\bigg( \Big( \sum_{i=0}^s \frac{(c\cdot s)^i}{i!}\cdot e^{-c\cdot s}\Big)^t \bigg)\ , 
\end{equation}
for $t=m/s$, which approaches $0$ for $m\to \infty$.

For the case with backup options we try to get an approximation for possible threshold densities.
Let $\tnb$ and $\tb$ be the loads ${n}/{m}$ that identify the transition from where there is a feasible orientation and where there is no feasible orientation of $G$ without and with the backup option respectively.
To get approximations for $\tnb$ and $\tb$ we study different ranges of load factors $[\cs,\ce]$.
Specifically, for all $c$ where $c=\cs+i\cdot 10^{-4}\leq c^{\mathrm{end}}$, and  $i=0,1,2,\ldots,$ we construct $a$ random graphs 
and measure the failure rate $\fr$ at $c$. We fit the sigmoid function 
\begin{equation}
 \label{eq:sigmoid}
f(c;x,y) = \big(1+\exp( -(c-x)/y ) \big)^{-1} 
\end{equation}
to the data points ($c$, $\fr$) using the method of least squares. The parameter $x$ (inflection point) is an approximation of $\tnb$ and $\tb$ respectively. With $\res$ we denote the sum of squares of the residuals.

Furthermore, for different $c$ and page sizes $s$, we are interested in the maximum ratio $\npn=\nprim/n$ or load $\npm=\nprim/m$ of primary keys, respectively.

For a fixed page $p$ let $\wb$ be the number of keys that have primary page $p$ but are inserted on their backup page.
Since the number of potential primary keys for a page follows a binomial distribution, 
some pages will be lightly loaded and therefore have a small value of $\wb$ or even $\wb=0$.
Some pages will be overloaded and have to shed load, yielding a large value of $\wb$.
We want to study the relative frequency of the values $\wb$.

\subsubsection{Results.}  Here we consider results from an optimal placement algorithm.

\begin{figure}
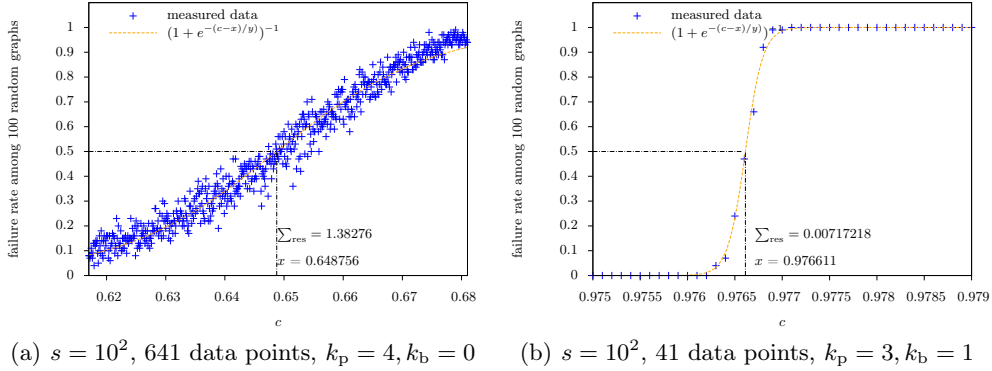

\subfigure[ $s=10^2$, $641$ data points, $\kprim=4, \kback=0$ ]{\scalebox{\imgScale}{\input{\figPath/maxCardMatch_m10E6_t10E4_dp4_ds0_a10E2_TEX.tex}}}
\subfigure[ $s=10^2$, $41$  data points, $\kprim=3, \kback=1$ ]{\scalebox{\imgScale}{\input{\figPath/maxCardMatch_m10E6_t10E4_dp3_ds1_a10E2_TEX.tex}}}
\caption{\label{fig:threshold_curves}Point of transition (a) without and (b) with backup pages.}
\end{figure}

\paragraph{I.}Table \ref{tab:thresholds} gives approximations of the loads where cuckoo hashing with paging and $k=4$ hash functions has failure rate $\fr=0.5$ in the case of $1$ or $0$ backup pages. With no backup pages the number of keys that can be stored decreases with decreasing page size and the success probability around $\tnb$ converges less rapidly, as demonstrated clearly in Fig.~\ref{fig:threshold_curves}. This effect becomes stronger as the pages get smaller. For this reason the range of load factors $[\cs,\ce]$ of sub-table~(a) grows with decreasing page size. Using only one backup edge per key almost eliminates this effect.
In this case the values $\tb$ seem to be stable for varying $s$ and are very near to the theoretical threshold of standard $4$-ary cuckoo hashing, which is $c^*_4\approx0.976770$; only in the case of very small pages $s=10$ can a minor shift of $\tb$ be observed. 
The position of $\tb$ as well as the slope of the fitting function appear to be quite stable for all considered page sizes. 

\newcommand{\ws}[1]{\hspace{1.2em}#1\hspace{1.2em}}
\begin{table}
\scalebox{0.90}{
\subtable[$40 \cdot 2^{6-\log_{10}(s)}+1$ data points, $\kprim=4, \kback=0$]{
\begin{tabular}{lccl}
 $s$      & \ws{$[\cs, \ce]$} & $\tnb$     & $\res$     \\\hline
 $10^6$   & $[0.975, 0.979]$ & 0.976794  & 0.014874  \\
 $10^5$   & $[0.968, 0.976]$ & 0.971982  & 0.096023 \\
 $10^4$   & $[0.944, 0.960]$ & 0.952213  & 0.299843  \\
 $10^3$   & $[0.863, 0.895]$ & 0.879309  & 0.653894  \\
 \rowcolor{lightlightgray}$10^2$   & $[0.617, 0.681]$ & 0.648756  & 1.382760   \\
 $10^1$   & $[0.124, 0.252]$ & 0.188029  & 1.809620   \\
\end{tabular}
}

\subtable[$41$ data points, $\kprim=3, \kback=1$]{
\begin{tabular}{lccl}
 $s$      & \ws{$[\cs,\ce]$} & $\tb$     & $\res$     \\\hline
\\
 $10^5$   & $[0.975, 0.979]$  &  0.976774 & 0.010523 \\  
 $10^4$   &      ''         &  0.976760 & 0.014250 \\
 $10^3$   &      ''         &  0.976765 & 0.002811 \\
 \rowcolor{lightlightgray}$10^2$   &      ''         &  0.976611 & 0.007172 \\
$10^1$   & $[0.9712, 0.9752]$ &   0.973178 & 0.008917 \\
\end{tabular}
}
}

\caption{\label{tab:thresholds}Approximations of the load factors that are the midpoints of the transition from failure rate 0 to failure rate 1 (without and with backup option) via fitting function
\eqref{eq:sigmoid} to a series of data points. For each data point the failure rate among $100$ random graphs was measured.
The grey rows correspond to the plots of Fig.~\ref{fig:threshold_curves}.}

\end{table}

\paragraph{II.}The average of the {maximum fraction of primary keys}, allowing one backup option, is shown in Table \ref{tab:opt_frac_primary}.
The fraction decreases with increasing load factor $c$ and decreases with decreasing page size $s$ as well.
Interestingly, for several parameters,
we found that an optimal algorithm finds placements with more than $c^*_3\cdot m$ keys sitting in one of their $3$ primary positions, where $c^*_3\approx0.917935$ is the threshold for standard $3$-ary cuckoo-hashing.
That is, more keys obtain one of their primary three choices with three primary and one backup choice than what could be reached using just three primary choices even without paging.   

\newcommand{\col}{\cellcolor{lightgray}}
\begin{table}[ht]
\vspace{0.3cm}
\scalebox{\tabScale}{
 \begin{tabular}{c||cc|cc|cc|cc|cc}
      & \multicolumn{2}{c|}{$s=10^5$} &\multicolumn{2}{c|}{$s=10^4$} &\multicolumn{2}{c|}{$s=10^3$} &\multicolumn{2}{c|}{$s=10^2$} &\multicolumn{2}{c}{$s=10^1$} \\
$c$  &  $\Enpn$  & $\Enpm$   &$\Enpn$    & $\Enpm$   &$\Enpn$    &$\Enpm$    &$\Enpn$    &$\Enpm$    &$\Enpn$    & $\Enpm$         \\\hline
0.90 & 1.000000 & 0.900000 & 0.999881 & 0.899893 & 0.995650 & 0.896085 & 0.975070 & 0.877563 & 0.902733 & 0.812460 \\
0.91 & 0.999997 & 0.909997 & 0.999093 & 0.909175 & 0.993008 & 0.903638 & 0.971556 & 0.884116 & 0.898281 & 0.817436 \\
0.92 & 0.998136 &\col 0.918286 & 0.996111 &     0.916422 & 0.989452 & 0.910296 & 0.967781 & 0.890358 & 0.893546 & 0.822062 \\
0.93 & 0.990957 &\col 0.921510 & 0.990467 &\col 0.921134 & 0.985015 & 0.916064 & 0.963723 & 0.896263 & 0.888041 & 0.825878 \\
0.94 & 0.983443 &\col 0.924436 & 0.983422 &\col 0.924416 & 0.979730 &\col 0.920946 & 0.959429 & 0.901863 & 0.880848 & 0.827997 \\
0.95 & 0.975952 &\col 0.927154 & 0.975961 &\col 0.927163 & 0.973744 &\col 0.925057 & 0.954876 & 0.907132 & 0.872427 & 0.828805 \\
0.96 & 0.968578 &\col 0.929835 & 0.968524 &\col 0.929783 & 0.967224 &\col 0.928535 & 0.947650 & 0.909744 & 0.862883 & 0.828367 \\
0.97 & 0.961112 &\col 0.932279 & 0.961157 &\col 0.932323 & 0.956892 &\col 0.928185 & 0.935928 & 0.907850 & 0.850154 & 0.824650 \\
 \end{tabular}
}
\vspace{0.2cm}            
\caption{\label{tab:opt_frac_primary}Average (among 100 random graphs) of the fraction of keys that can be placed on their primary page for different page sizes $s$ and $\kprim=3$, $\kback=1$. The failure rate is $\fr=0$. For $c\ge0.98$ the random graph did not admit a solution anymore. The entries
of the grey cells are larger than $c^*_3$.}
\end{table}               

\begin{figure}
\centering
\scalebox{\imgScale}{\input{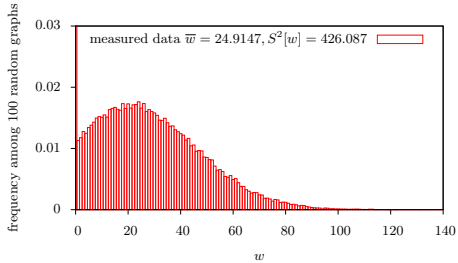}}
\vspace{-0.1cm}
\caption{\label{fig:moved_keys_per_page}frequency of $\wb=0$ is $0.169$, $(c,s)=(0.95,10^3)$, $a=10^3$, $\fr=0$}
\end{figure}

\paragraph{III.}
Figure~\ref{fig:moved_keys_per_page} depicts the relative frequency of the values $\wb$ among $10^5$ pages for selected parameters $(c,s)=(0.95,10^3)$. In this case about 17 percent of all pages do not need backup pages, \ie, $w=0$. This is consistent with the idea that pages with a load below $c^*_3 \cdot s$ will generally not need backup pages. The mean $\Ewb$ is about 2.5 percent of the page size $s$ and for about $87.6$ percent of the pages the value $\wb$ is at most $5$ percent of the page size. The relative frequency of $\wb$ being greater than $0.1s$ is very small, about $1.1\cdot 10^{-3}$.
 
\subsubsection{Summary.}
\label{sec:page_request}
We observed that using pages with $(\kprim,\kback)=(3,1)$ we achieve loads very close to the $c^*_4$ threshold ($\tb\approx c^*_4$).
Moreover the load $\npm$ from keys placed on their primary page $\npm$ is quite large, near or even
above $c^*_3$.

Let $X$ be the average (over all keys that have been inserted) number of page requests needed in a search for a key $x$,
where naturally we first check the primary page.
If $(\kprim,\kback)=(3,1)$ and a key was equally likely to be in any of its locations, 
the \emph{expected number of page requests} $\mathrm{E}(X)$ would satisfy $\mathrm{E}(X)=1.25$.
If $(\kprim,\kback)=(3,1)$ and $c$ is near $c^*_4$ then we have roughly $\mathrm{E}(X)\approx c^*_3/c \cdot 1 + (1-c^*_3/c) \cdot 2$.
For example, for $(c,s)=(0.95,10^3)$, using the values of Table \ref{tab:thresholds} we find $\mathrm{E}(X)\approx 0.974 \cdot 1 +  0.026\cdot 2<1.03$.

Now assume we perform a lookup for a key $x$ not in the table. The
disadvantage of using two pages per key is that now we always require two
page requests, \ie, $\mathrm{E}(X)=2$. This can be circumvented by
storing an additional set membership data structure, such as a Bloom
filter \cite{Bloom1970}, for each page $p$ representing the $\wb$ many
keys that have primary page $p$ but are inserted on their backup page.

One can trade off space, computation, and the false positive
probability of the Bloom filter as desired.  As an example, suppose
the Bloom filters use $3$ hash functions and their size corresponds to
just one bit per page cell. In this case, we can in fact use the same
hash functions that map keys to cell locations for our Bloom filter.
Bounding the fraction of 1 bits of a Bloom Filter
from above via $(\kprim \cdot w)/s$, the distribution of $\wb$ as in
Fig.~\ref{fig:moved_keys_per_page} leads to an average false positive
rate of less than $0.15$ percent and therefore an expected number of page
requests $E(X)$ of less than $1.0015$ for unsuccessful searches.  One could
reduce false positives even further using more hash functions, or use
less space.

\subsection{Dynamic Case}
We have seen the effectiveness of optimal offline cuckoo hashing with paging.
We now investigate whether similar placements can be found online, by considering the simple random walk algorithm
from Sect.~\ref{sec:randomWalk}.  We begin with the case of insertions only.

\subsubsection{Setup and Measurements.}
Along with the failure rate $\fr$, 
the fraction of primary keys $\npn$ and corresponding load $\npm$, 
and the distribution of the number of keys $\wb$ inserted on their backup page,
we consider two more performance characteristics:
\begin{compactitem}
\item[$\st$] - the average number of steps of the random walk insertion procedure. A step is either storing a key $x$ in a free cell $y$ or replacing an already stored key with the current ``nestless'' key.
\item[$\pr$] - the average number of page requests over all inserted items. Here each new key $x$ requires at least one page request, and every time we move an item to its backup page, that requires another page request. 
\end{compactitem}

We focus on characteristics of the algorithm with loads near $\tb$, varying the number of table cells $m=10^5,10^6,10^7$ and page sizes $s=10,10^2,10^3$. The performance of the algorithm heavily depends on the choice of parameters $\palpha$ and $\pbeta$. Instead of covering the complete parameter space we first set $\pbeta$ to infinity and use the measurements to give insight into the performance of the algorithm for selected values of $\palpha$.

In addition we want to explore whether we can expect a sufficiently low failure probability of the random walk algorithm, at least for some selected sets of parameters $\pi$ including practical values of $\pbeta$. For this we tested the following null hypothesis 
$H_0(\pi)=$``If one uses parameter set $\pi$ then Algorithm \ref{algo:random_walk} fails with probability at least $p$.''
To test the null hypothesis for a specific $\pi$ we performed the random experiment ``insertion of $n=c\cdot m$ keys with Algorithm \ref{algo:random_walk}'' $a$ times.
Let $A(\pi)$ be the event that all of the $a$ many random experiments for a given $\pi$ ended successfully. Then we have:
\begin{equation}
 \Pr\big( A(\pi) \mid H_0(\pi)\big)\leq (1-p)^a \leq \exp(-p\cdot a) \ .
\end{equation}
For example if $a=10^6$ and $p=10^{-5}$ we have $\Pr( A(\pi) \mid H_0(\pi))\leq \exp(-10)\approx 4.54 \cdot 10^{-5}$. 
Hence if we observe $A(\pi)$ we may reject the null hypothesis with high confidence.

We also study the influence of $\palpha$ for a fixed configuration.
We vary $\palpha$ to see qualitatively how the number of primary
keys as well as the number of steps and page requests depend on this
parameter.

It is well known that hashing schemes can perform differently in
settings with insertions and deletions rather than insertions alone,
so we investigate whether there are substantial differences in this
setting.  Specifically, we consider the table under a constant load by
alternating insertion and deletion steps.

\subsubsection{Results.}  Here we consider results from the random walk algorithm.
\paragraph{I.} Tables \ref{tab:random_walk_0.95} and \ref{tab:random_walk_0.97} show the behavior of the random walk algorithm with loads near $\tb$ for $(c,\palpha)=(0.95,0.97)$ and $(c,\palpha)=(0.97,0.90)$. The number of allowed steps for the insertion of $n$ keys is set to infinity via $\pbeta=\infty$. The number of trials $a$ per configuration is chosen such that $a\cdot m=10^9$ (keeping the running time for each configuration approximately constant).

We first note that with these parameters the algorithm found a placement for the keys in all experiments; failure did not occur.  For fixed page size the sample means are almost constant; for growing page size the load $\Enpm$ increases, while $\Est$ and $\Epr$ decrease, with a significant drop from page size $10$ to $100$.
For our choices of $\palpha$ the random walk insertion procedure missed the maximum fraction of primary keys by up to $2$ percent for $c=0.95$ and by up to $6$ percent for $c=0.97$ and needs roughly the same average number of steps (for fixed page size).

\begin{table}[!hb]
\centering
\scalebox{0.90}{
\begin{tabular}{c|ccccccccc}
$s$ & $m$    & $a$    & $t$       & $\Enpn$  & $\Enpm$  & $\Est$     & $\Vst$     & $\Epr$    & $\Vpr$   \\ \hline\hline
$10^1$ & $10^5$ & $10^4$ & $10^4$ & 0.860248 & 0.817236 & 158.707669 & 114.072760 & 10.327258 & 0.409334 \\
 ''    & $10^6$ & $10^3$ & $10^5$ & 0.860219 & 0.817208 & 158.618752 &  11.405092 & 10.321981 & 0.040869 \\
 ''    & $10^7$ & $10^2$ & $10^6$ & 0.860217 & 0.817206 & 158.645056 &   1.092781 & 10.323417 & 0.003914 \\ \hline
$10^2$ & $10^5$ & $10^4$ & $10^3$ & 0.938431 & 0.891509 & 22.807328 & 1.081478 & 2.248953 & 0.003760 \\
 ''    & $10^6$ & $10^3$ & $10^4$ & 0.938424 & 0.891503 & 22.813986 & 0.104012 & 2.249273 & 0.000366 \\
 ''    & $10^7$ & $10^2$ & $10^5$ & 0.938412 & 0.891491 & 22.813905 & 0.010862 & 2.249201 & 0.000038 \\ \hline
$10^3$ & $10^5$ & $10^4$ & $10^2$ & 0.955773 & 0.907985 & 16.580150 & 0.512018 & 1.892190 & 0.001779\\
 ''    & $10^6$ & $10^3$ & $10^3$ & 0.955737 & 0.907950 & 16.603145 & 0.052386 & 1.893515 & 0.000182\\
 ''    & $10^7$ & $10^2$ & $10^4$ & 0.955730 & 0.907943 & 16.598381 & 0.005534 & 1.893248 & 0.000019\\
\end{tabular}
}
\vspace{0.3cm}
\caption{\label{tab:random_walk_0.95}Characteristics of Algorithm \ref{algo:random_walk} for $(c,\palpha,\pbeta)=(0.95,0.97,\infty)$.  $\fr=0$.}
\end{table}

\begin{table}[!ht]
\centering
\scalebox{0.90}{
\begin{tabular}{c|cccccccccc}
$s$    & $m$    & $a$    & $t$    &  $\Enpn$ &  $\Enpm$  & $\Est$     & $\Vst$      & $\Epr$    & $\Vpr$   \\\hline\hline
$10^1$ & $10^5$ & $10^4$ & $10^4$  & 0.816795 & 0.792291 & 158.506335 & 1222.640379 & 32.336112 & 48.892079  \\
''     & $10^6$ & $10^3$ & $10^5$  & 0.816790 & 0.792286 & 153.645339 &   78.581917 & 31.363876 &  3.142566  \\
''     & $10^7$ & $10^2$ & $10^6$  & 0.816802 & 0.792298 & 152.873602 &   10.759338 & 31.209210 &  0.430827  \\\hline
$10^2$ & $10^5$ & $10^4$ & $10^3$ & 0.886997 & 0.860387  & 23.320507 & 2.731285 & 5.361922 & 0.108700\\
''     & $10^6$ & $10^3$ & $10^4$ & 0.886992 & 0.860382  & 23.289233 & 0.256942 & 5.355625 & 0.010218\\
''     & $10^7$ & $10^2$ & $10^5$ & 0.886985 & 0.860375  & 23.268641 & 0.024796 & 5.351518 & 0.000986\\\hline
$10^3$ & $10^5$ & $10^4$ & $10^2$ & 0.898281 & 0.871332  & 19.497032 & 1.550490 & 4.607751 & 0.061739 \\
''     & $10^6$ & $10^3$ & $10^3$ & 0.898232 & 0.871285  & 19.486312 & 0.146267 & 4.605481 & 0.005816 \\
''     & $10^7$ & $10^2$ & $10^4$ & 0.898235 & 0.871288  & 19.493215 & 0.012744 & 4.606893 & 0.000507 \\
\end{tabular}
}
\vspace{0.3cm}
\caption{\label{tab:random_walk_0.97}Characteristics of Algorithm \ref{algo:random_walk} for $(c,\palpha,\pbeta)=(0.97,0.90,\infty)$.  $\fr=0$.}
\end{table}

\paragraph{II.}To get more practical values for $\pbeta$ we scaled up the values $\Est$ from Tables \ref{tab:random_walk_0.95} and \ref{tab:random_walk_0.97} and {estimated the failure probability} for suitable parameter sets $\pi=(c,s,\palpha,\pbeta)\in\{
{\scriptstyle(0.95, 10^2, 0.97, 30 ), (0.95, 10^3, 0.97, 25 ), (0.97, 10^2, 0.90, 30 ), (0.97, 10^3, 0.90, 25 )}\}$.
For all these parameter sets we observed a failure rate of zero among $a=10^6$ attempts (event $A(\pi)$). We can conclude at a level of significance of at least $1-e^{-10}$ that for these sets the failure probability of the random walk algorithm is at most $10^{-5}$.

\paragraph{III.}Figure \ref{fig:parameter_alpha} shows how parameter $\palpha$ influences the 
ratio of primary keys $\npn$, the number of insertion steps $\st$ and the number of page requests $\pr$.

\begin{figure}[htb]
\centering
\subfigure{\scalebox{0.58}{\input{\figPath/randWalk_m10E6_t10E3_dp3_ds1_a10E3_fac30_thresh070-099_loadTEX.tex}}}
 \hspace{0.2cm}
\subfigure{\scalebox{0.58}{\input{\figPath/randWalk_m10E6_t10E3_dp3_ds1_a10E3_fac30_thresh070-099_stepTEX.tex}}}
 \hspace{0.2cm}
\subfigure{\scalebox{0.58}{\input{\figPath/randWalk_m10E6_t10E3_dp3_ds1_a10E3_fac30_thresh070-099_preqTEX.tex}}}
\caption{\label{fig:parameter_alpha} $(c,s,\pbeta)=(0.95,10^3,30), a=10^3, \fr=0$}
\end{figure}

The mean fraction of primary keys $\Enpn$ grows linearly and $\Est$ grows nonlinearly with growing $\palpha$.
For $\palpha=0.98$ the gap between the optimal fraction of primary keys and the fraction reached by the random walk procedure is about $1$ percent. The value of $\Est$ also depends nonlinearly on $\palpha$ and reaches a local minimum at $\palpha=0.95$. The sample variances are quite small and stable except for $S[\st]$ and large $\palpha$ (near $0.98$).

\begin{figure}[ht]
\centering
{
\scalebox{0.7}{
\input{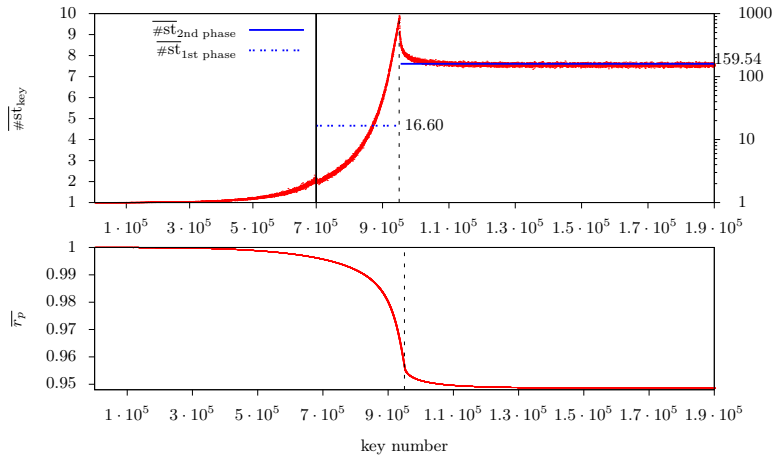}
}
\caption{\label{fig:number_of_steps}$(c,s,\palpha,\pbeta)=(0.95,10^3,0.97,30)$, $a=10^3$, $\fr=0$, The ordinate of the right half of the upper plot is in log scale.}
}
\end{figure}

\paragraph{IV.}
The results for alternating insertions and deletions for parameters $(c,s)=(0.95,10^3)$ and $(\palpha,\pbeta)=(0.97,30)$ are shown in Fig.~\ref{fig:number_of_steps}. We measured the current fraction of primary keys $\npn$ and the number of insertion steps with respect to each key $\st_{\mathrm{key}}$. Recall that $\st$ is the average number of insertion steps concerning all keys.
In the first phase (insertions only) the average number of steps per key grows very slowly at the beginning and is below $10$ when reaching a load where about $1$ percent of current keys are backup keys.  After that $\Est_{\mathrm{key}}$ grows very fast up to almost $10^3$ (for the last few keys), which is the page size.
The sample mean of the average number of steps $\Est$ up to this point is about $16.6$. Similarly the sample mean of the fraction of primary keys $\Enpn$ decreases very slowly at the beginning and decreases faster at the end of the first phase.
Up to load about $82.6$ percent the fraction of backup keys is below $1$ percent. 
In the second phase (deletions and insertions alternate) $\Est_{\mathrm{key}}$ and $\Est$ decrease and quickly reach a steady state. Since the decrease of $\Enpn$ is marginal but the drop $\Est_{\mathrm{key}}$ is significant we may conclude that the overall behavior is better in steady state than at the end of the insertion only phase. Moreover in an extended experiment with $n=c\cdot m$ insertions and $10\cdot n$ delete-insert pairs the observed equilibrium remains the same and therefore underpins the conjecture that Fig.~\ref{fig:number_of_steps} really shows a ``convergence point'' for alternating deletions and insertions.

\paragraph{V.}
Figure~\ref{fig:moved_keys_per_page_random_walk} shows the relative frequency of the values $\wb$ among $10^5$ pages for $(c,s)=(0.95,10^3)$ and $(\palpha,\pbeta)=(0.97,30)$ at the end of the insertion only phase, given by Fig.~\ref{fig:moved_keys_per_page_random_walk}~(a), and
at the end of the alternation phase, given by Fig.~\ref{fig:moved_keys_per_page_random_walk}~(b).
\begin{figure}
\centering
\hspace{0.7cm}
\subfigure[insertion only phase]{\scalebox{\imgScale}{\input{\figPath/randWalk_m10E6_t10E3_dp3_ds1_a10E2_fac30_thresh097_backup_per_pageTEX.tex}}}
 \hspace{0.9cm}
\subfigure[insertion and deletion phase]{\scalebox{\imgScale}{\input{\figPath/del_2n_randWalk_m10E6_t10E3_dp3_ds1_a10E2_fac30_thresh097_backup_per_pageTEX.tex}}}
\caption{\label{fig:moved_keys_per_page_random_walk}frequency of $\wb$, $(c,s)=(0.95,10^3)$, $a=10^3$, $\fr=0$}
\end{figure}
Note that Fig.~\ref{fig:moved_keys_per_page_random_walk}~(a) corresponds to Fig.~\ref{fig:moved_keys_per_page} with respect to the graph parameters. 
The shapes of the distributions differ only slightly, except that in the second phase the number of backup keys is larger.
In comparison with the values given by the optimal algorithm in Fig.~\ref{fig:moved_keys_per_page} the distribution of the
$\wb$ values is more skewed and shifted to the right.

\subsubsection{Summary.}
A simple online random-walk algorithm, with appropriately chosen parameters, can perform quite close to the optimal algorithm for cuckoo hashing with paging, even in settings where deletions occur.  

With parameters $(c,s)=(0.95,10^3)$ and $(\palpha,\pbeta)=(0.97,30)$ the expected number
of page requests $E(X)$ for a successful search is about $1.044$, using the values from
Table \ref{tab:random_walk_0.95}.
With the Bloom filter approach described in Sect.~\ref{sec:page_request} (which can be done only after finishing the insertion of all keys), the distribution from Fig.~\ref{fig:moved_keys_per_page_random_walk}~(a) gives an expected number of page requests for an unsuccessful search of less than $1.0043$.
Both values are only slightly higher than those resulting from an optimal solution.  One can instead use counting Bloom filters \cite{broder2004network} to improve performance for unsuccessful searches with online insertions and deletions, at the cost of more space.

\subsection{Small Pages}
We have seen that if one uses one backup option then the page size has only marginal influence on the existence of a legal orientation of $G$ but heavily influences the maximum fraction of primary keys (in the dynamic case as well as in the static case). 
Tables \ref{tab:opt_frac_primary}, \ref{tab:random_walk_0.95} and \ref{tab:random_walk_0.97} show that the smaller the page the 
smaller the fraction of primary keys, with a significant decrease from page size $100$ to $10$.
In order to attenuate this downside one can use the following variant.
Let $\kprim=1$ and $\kback=1$. We use the idea of blocked cuckoo hashing \cite{CainSW2007,dietzfelbinger2007balanced,FernholzR2007}
where each table cell gets capacity $\ell$ for some constant $\ell$. One can think of pages of size exactly one cell. Some reference (theoretical) thresholds $c^*_{2,\ell}$ for the existence of a legal orientation of the corresponding cuckoo graphs are given in Table \ref{tab:2,l-ary} \cite{CainSW2007,FernholzR2007}; and experimental threshold values are given in Figure \ref{fig:2,l-ary}. (Note that $m$ is the number of table cells of capacity $\ell$ and we refer to the normalized values $c^*_{2,\ell}/\ell$.)

\begin{table}[htb]
\centering
\begin{tabular}{c|ccccccc}
 $\ell$                   &  2        & 3        & 4        & 5        & 8        & 10       & 16    \\\hline
${c^*_{2,\ell}}/{ \ell }$ &  0.897012 & 0.959154 & 0.980370 & 0.989551 & 0.997853 & 0.999143 & 0.999928
\end{tabular}
\vspace{0.2cm}
\caption{\label{tab:2,l-ary}Theoretical thresholds values $c^*_{2,\ell}$.}
\end{table}

%

\begin{figure}
\subfigure[$\ell=4$]{\scalebox{\imgScale}{\input{\figPath/maxCardMatch_m10E6_t10E0_dp4_ds4_a10E2_b=4_TEX.tex}}}
\subfigure[$\ell=16$]{\scalebox{\imgScale}{\input{\figPath/maxCardMatch_m10E6_t10E0_dp16_ds16_a10E2_b16_TEX.tex}}}
\caption{\label{fig:2,l-ary} $m=10^6/\ell$, $\kprim=1, \kback=1$,  $a=10^2$}
\end{figure}

Our aim remains to store as many keys as possible in their primary cell while keeping the load $c$ near the threshold $c^*_{2,\ell}$. 
Table \ref{tab:2,l-ary_opt_prim} gives optimal (offline) results.
\begin{table}
\vspace{0.3cm}
\centering
\scalebox{0.9}{
\begin{tabular}{c||cc|cc|cc|cc}
     & \multicolumn{2}{c|}{$\ell=4$} &\multicolumn{2}{c|}{$\ell=8$} &\multicolumn{2}{c}{$\ell=10$}&\multicolumn{2}{c}{$\ell=16$} \\
$c/\ell$ &  $\Enpn$  & $\Enpm/\ell$ &  $\Enpn$  & $\Enpm/\ell$   &$\Enpn$    & $\Enpm/\ell$   &$\Enpn$    &$\Enpm/\ell$  \\\hline
0.90 & 0.822251 & 0.740026    & 0.898282 & 0.808454 &    0.913798 & 0.822418      & 0.940022 & 0.846020\\
0.91 & 0.815652 & 0.742244    & 0.894259 & 0.813776 &    0.910014 & 0.828113      & 0.936509 & 0.852223\\
0.92 & 0.808867 & 0.744157    & 0.890040 & 0.818837 &    0.906196 & 0.833700      & 0.932937 & 0.858302\\
0.93 & 0.801424 & 0.745325    & 0.885679 & 0.823682 &    0.902177 & 0.839024      & 0.929188 & 0.864145\\
0.94 & 0.793452 & 0.745845    & 0.881098 & 0.828232 &    0.898052 & 0.844169      & 0.925333 & 0.869813\\
0.95 & 0.784526 & 0.745300    & 0.876222 & 0.832411 &    0.893687 & 0.849003      & 0.921360 & 0.875292\\
0.96 & 0.774254 & 0.743283    & 0.870778 & 0.835947 &    0.889150 & 0.853584      & 0.917347 & 0.880653\\
0.97 & 0.761745 & 0.738893    & 0.864615 & 0.838676 &    0.884317 & 0.857787      & 0.913244 & 0.885846\\
0.98 & 0.743799 & 0.728923    & 0.857017 & 0.839876 &    0.878957 & 0.861378      & 0.908929 & 0.890751\\
0.99 & \multicolumn{2}{c|}{no solution}          & 0.847632 & 0.839156 &    0.870738 & 0.862030      & 0.904474 & 0.895429\\
\end{tabular}
}
\vspace{0.2cm}
\caption{\label{tab:2,l-ary_opt_prim}Maximum fraction of primary keys among 100 random graphs, for different block sizes $\ell$, $m=10^6/\ell$. For $c/\ell=0.98,\ell=4$ the failure rate is $\fr=0.01$ and for $c/\ell=0.99,\ell=4$ we have $\fr=1$; otherwise $\fr=0$.  
}
\end{table}
They indicate that with respect to the ratio of primary keys the variant $(\kprim,\kback,s,\ell)=(1,1,1,10)$
is slightly better than  $(\kprim,\kback,s,\ell)=(3,1,10,1)$. An advantage is that for $\ell>3$ the (known) thresholds $c^*_{2,\ell}/\ell$ are higher than the values $\tb$ which are near $c_4^*$ $(c_4^*\approx 0.976770)$.
For example, with $\ell=16$ and load factor $c/\ell=0.99$ a fraction $0.904$ of the keys can be stored in their
primary cell, thus reducing the expected number of cell requests for successful searches from $1.5$
when each key is equally likely to be in either location to less than $1.1$. 

\section{Conclusion}
Our results suggest that cuckoo hashing with paging may prove useful in a number of settings where the cost of multiple lookups might otherwise prove prohibitive. Perhaps the most interesting aspect for continuing work is to obtain
provable performance bounds for cuckoo hashing with pages.  Even in
the case of offline key distribution with one additional choice on a
second page we do not have a formal result proving the threshold
behavior we see in experiments.     


\end{document}